# SwinCCIR: An end-to-end deep network for Compton camera imaging reconstruction

Minghao Dong, *Graduate Student Member*, *IEEE*, Xinyang Luo, Xujian Ouyang, Yongshun Xiao

*Abstract*—**Compton cameras (CCs) are a kind of gamma cameras which are designed to determine the directions of incident gammas based on the Compton scatter. However, the reconstruction of CCs face problems of severe artifacts and deformation due to the fundamental reconstruction principle of back-projection of Compton cones. Besides, a part of systematic errors originated from the performance of devices are hard to remove through calibration, leading to deterioration of imaging quality. Iterative algorithms and deep-learning based methods have been widely used to improve reconstruction. But most of them are optimization based on the results of back-projection. Therefore, we proposed an end-to-end deep learning framework, SwinCCIR, for CC imaging. Through adopting swin-transformer blocks and a transposed convolution-based image generation module, we established the relationship between the list-mode events and the radioactive source distribution. SwinCCIR was trained and validated on both simulated and practical dataset. The experimental results indicate that SwinCCIR effectively overcomes problems of conventional CC imaging, which are expected to be implemented in practical applications.**

## I. INTRODUCTION

COMPTON cameras[1] (CCs) are a kind of gamma cameras which utilize the quantitative relationship between the energy of the scattered photon and the scatter angle in the Compton scatter to determine the incident direction of gammas. Because of the different imaging principle, CCs have the advantages of a wider field of view[2] and a higher applicable energy range compared to conventional gamma cameras and therefore have attracted considerable attention for potential applications in environmental security[3,4], medical diagnosis and treatment[5,6,7], space exploration[8,9,10,11], et.al.

CC reconstruction is usually based on list-mode data, which consists of a list of multi-interaction events. The CC record deposited energies and interaction positions of gammas and divided them into different events based on the interaction time. One effective multi-interaction event for CC reconstruction refers to a set of energy-position pairs recorded by the CC representing the Compton scattering or photoelectric absorption of the gamma and its secondary particles. Through the quantitative relationship between the scattering angle and the energy of scattered photons in Compton scatter, the incident direction of the gamma in a multi-interaction event can be constrained to a conical surface, which we also called the Compton cone.

However, limited by the time resolution of electronic devices, current CCs can hardly distinguish the order of interactions in one multi-interaction event, but simply treat interactions occurring in a preset time window as originating from the same gamma, which may also result in wrong coincidence. Additionally, the inherent energy resolution and position resolution of detectors will also lead to a considerable deviation in scatter angles[12]. Therefore, CC imaging severely suffered from artifacts and deformations.

To improve the image quality, optimization methods from different perspectives have been applied in CC reconstruction. On the one hand, iterative algorithms such as list-mode maximum likelihood expectation maximization[13,14,15] (MLEM) and stochastic origin ensemble[16] (SOE) adjust the initial distribution of radioactive sources, which usually is the result of simple back-projection (SBP), through certain criterions to make it gradually close to the distribution that is most likely to obtain the observed data, effectively suppressing artifacts and improving signal-to-noise ratio. One the other hand, event selection and interaction sorting methods based on energies[17, 18], polarization[19] and machine learning[20], et.al, have been thoroughly researched to improve the accuracy of scatter angles and reduce the negative impact of wrong coincidence events from the perspective of list-mode. Recently, deep learning methods[21,22] have also been applied to the above two tasks. The convolutional neural network (CNN) has been applied to determine the scattering position and electron recoil direction in a time projection chamber (TPC) CC from the raw data[23] and has also been used to predict the accurate source position based on initial SBP results[24]. Yao et.al[25] applied a U-net to achieve resolution recovery of the SBP results. However, most of these researches optimized the CC reconstruction in a single domain, either in list event data or in initial reconstructed images. Though these models directly improve the quality of CC imaging, they cannot "learn" or eliminate part of systematic errors, especially those from the detectors. For the models applied to the data domain, the ground truth of list data would be inevitably affected by systematic errors. Similarly, the input of models applied in image domain would also be influenced due to the inaccuracy in data and labels. Both problems will lead to overfitting in practical applications, which is also one of the reasons why most researches are based on Monte Carlo (MC) simulation data rather than

(Corresponding author: Yongshun Xiao, xiaoysh@mail.tsinghua.edu.cn)
Minghao Dong, Xinyang Luo, Xujian Ouyang and Yongshun Xiao are all with the Department of Engineering Physics, Tsinghua University, Beijing, 100084, China (e-mail: dongmh22@126.com; luo-xy22@mails.tsinghua.edu.cn; oyxj21@mails.tsinghua.edu.cn; xiaoysh@mail.tsinghua.edu.cn).

practical experiments.

To overcome the discontinuity between the data and image domains, a recent study has proposed a direct CC reconstruction network, the ComptonNet[26], based on multilayer perceptron (MLP) to directly connect the list data domain and the image domain and achieve better results compared to some conventional reconstruction algorithms. Pitifully, this study is still based on simulation data where practical systematic errors such as the position errors of sources and detectors have not been fully considered. Besides, the input of ComptonNet contains a large amount of non-Compton scatter events, which might enhance the imaging performance but also limits the number of effective events and the universality of the model. To address these issues, we propose a swin-transformer-based end-to-end network for CC image reconstruction (SwinCCIR) to directly map list-mode data to images and tested on real experimental data. There are two major differences between our model with the ComptonNet. First, we input only multi-interaction events to map more direct CC imaging. Second, we use swin-transformer block[27] to extract features determined by multiple events, which is consistent with the basic CC reconstruction principle. Thus, we expect such design can better establish the mapping between the data domain and the image domain.

## II. BACKGROUND

### A. The basic reconstruction principle

The fundamental physical principle relied upon by CC imaging is the Compton scatter. In the Compton scatter, a gamma photon interacts with an extranuclear electron and changes its direction and energy, which is similar to the elastic collision in macroscopic physics. Due to the adherence to the conservation of momentum and energy, the energy of the scattered photons corresponds one-to-one with the scattering angle. The CC records the deposited energy and position for each interaction and therefore can determine the incident direction of the gamma photon. A simplest case of the event expected to be captured by the CC is the ideal two-interactions gamma event, where the incident gamma photon undergoes Compton scatter and then is absorbed. If the energy loss of the photon in Compton scatter is $E_1$ and the energy of the scattered photon is $E_2$, the scatter angle $\theta$ can be determined through (1). As the Fig.1(a) shows, the incident direction of the photon scattered at $P_1$ and was absorbed at $P_2$ can be constrained to a conical surface with $P_1$ as the apex, $\overrightarrow{P_2P_1}$ as the axis, and a half-aperture angle of $\theta$, which we also called the Compton cone. By back-projecting Compton cones into the imaging space, the position or the orientation of the radioactive source will be highlighted as the Fig.1(b) shows. For events involve $N$ interactions and deposit all energies, scatter angles can be obtained by simply replacing $E_2$ in (1) with $\sum_{n=2}^{N} E_n$. The vector $\overrightarrow{P_2P_1}$ pointing from the second interaction point to the first one is commonly referred as the Compton lever.

$$\cos(\theta) = 1 - \frac{m_0 c^2 E_1}{E_2(E_1 + E_2)} \tag{1}$$

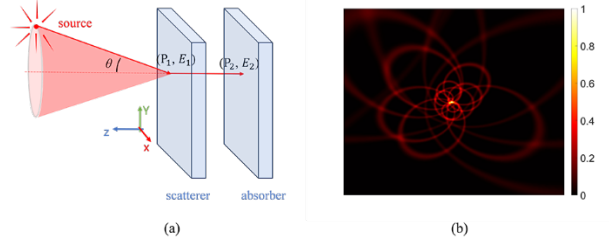

Fig. 1. (a) is the basic reconstruction schematic diagram. (b) is an example of CC reconstruction of a point-like source by simply back-projection the Compton cones.

However, current detectors (whose typical time resolution is $2.5 \times 10^2$ ps[28, 29]) cannot distinguish the interaction sequence due to the short interaction time interval ($\sim 100$ ps). Besides, there are cases where photons do not fully deposit energy, such as leaving the detector after scatter. The order of interaction and the initial energy of the photon need to be determined through some auxiliary methods with prior constraints, such as the energy relationship in Compton scatter and the spectral reconstruction algorithm.

Based on the reconstruction principle above, the CC reconstruction can be divided into two steps. One is the mapping from list-mode event data to Compton levers and scatter angles, and the other is image reconstruction and enhancement. Conversely, CC imaging system can be mathematically formulated as (2) and (3).

$$N(Y) \sim Possion(\mathbf{T}X) \tag{2}$$

$$N(\tilde{Y}) \sim D(N(Y)) \tag{3}$$

Where $X$ is the source property in the pixelated imaging space and $N(Y)$ is the number of Compton event $Y$ that occur. Generally, each $X_j \in X$ is the radiation intensity of the imaging space pixel $p_j$. Each $Y_i \in Y$ consists of $M_i$ position-energy pairs $(P_i, E_i)_{m=1,2\cdots M_i}$ arranged in the order of interactions. $\mathbf{T}$ is the probability matrix whose element $t_{ij}$ corresponds to the probability that photons emitted from $p_j$ undergo interactions in the form of $Y_i$ in the CC. $Y_i$ cannot be directly obtained due the limitation of the detector, i.e., the limitation of time resolution and energy resolution. Thus, we introduced an implicit probability distribution, $D(\cdot)$, to denote the influence of the detector. Physically, $D(\cdot)$ represents the signal generation process of the detectors, which is comprehensively affected by the performance and the design of the detector. $\tilde{Y}$ is the measured list-mode data where the order of position-energy pairs is given according to the spatial position of detectors and their values are also influenced by measurement uncertainties.

As discussed above, the CC imaging system can acquire $\tilde{Y}$ and the reconstruction aims to recover an estimate of the source distribution, $\hat{X}$, of $X$ from $\tilde{Y}$. Supposing the reconstruction is parameterized by $\theta$ and the operator is represented by $F_\theta(\cdot)$, the CC reconstruction can be expressed as the optimization (4).

$$\theta^* = \arg\min_\theta \mathcal{L}(\hat{X}, X) = \arg\min_\theta \mathcal{L}(F_\theta(\tilde{Y}), X) \quad (4)$$

Where $\mathcal{L}(\cdot, \cdot)$ is the loss function. With $F_\theta(\cdot)$ in the form of a deep network model, we obtain $\theta^*$ via supervised training with $(\tilde{Y}, X)$ pairs.

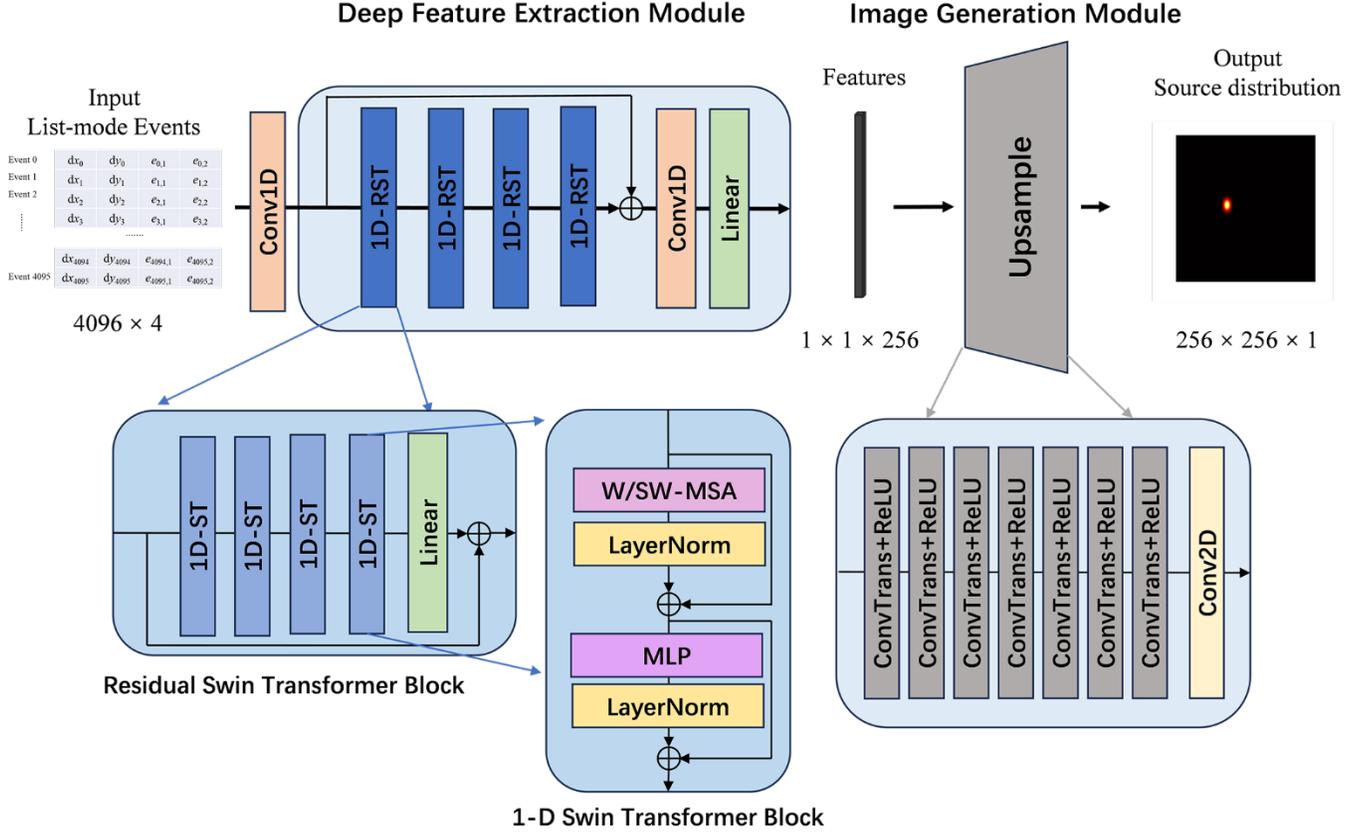

Fig. 2. The overall SwinCCIR architecture along with 1D-RST and 1D-ST subcomponents.

## B. The structure of the tested CC

CCs have many different structures and namely can be divided into single-layer type and multi-layer type. Single-layer CCs require the ability to distinguish the positions and energies of multiple interactions within a single detector, while multi-layer CCs usually record different interactions in different layers. The CC used in this study is a two-layer CC which consists of two scintillator detector units. The first layer[30], i.e., scatterer, consists of a $10 \times 10$ LaBr$_3$ array and 8 × 8 Sensl J60035 Silicon Photomultiplier (SiPM) array, each LaBr$_3$ pixel measures $5 \times 5 \times 5$ mm$^3$ and each SiPM measures $6 \times 6$ mm$^2$. The second layer[31], i.e., absorber, consists of a $25 \times 25$ GAGG array and the same SiPM array as the scatterer. The GAGG pixel measures $2 \times 2 \times 5$ mm$^3$. The distance between the two layers is 50 mm. Average energy resolutions of the scatterer and absorber are 7.64%@511 keV and 7.80%@511 keV, respectively. All scintillator pixels are coated with BaSO$_4$ reflector except for the light emitting surface. After the interaction of the gamma, the hit scintillator pixel emits scintillation photons that reflect inside the pixel and leave through the light emitting surface. The SiPM array attached to the light emitting surface absorb the scintillation photons and output 64-channels charge signals, which are further encoded into 4-channels signals. Using the centroid positioning algorithm, the interaction position can be approximately determined and resolved to pixel level, but the exact position within the pixel is not available. Furthermore, the interaction position would be calculated as the centroid weighted by deposited energies if the photon undergoes more than one interaction in a single layer. In this case, $\tilde{Y}_i$ in this study only consists of 4 key attributes $\{P_{i,1}, E_{i,1}, P_{i,2}, E_{i,2}\}$. Furthermore, this study focuses on determination of the orientation of distant radioactive sources, where the source-to-detector distance is much larger than the detector dimensions. In this case, all $P_{i,1}$ can be regarded as a single point, $\tilde{Y}_i$ then can be further simplified to $\{dx_i, dy_i, E_{i,1}, E_{i,2}\}$, where $dx_i$ and $dy_i$ are the components of $\overrightarrow{P_{i,2}P_{i,1}}$ in the x and y directions, $dz_i$ is omitted as it is a constant 50.

## III. THE PROPOSED METHOD

### A. Network Architecture

We propose the SwinCCIR whose architecture is illustrated in Fig. 2 to implement the direct reconstruction operator $F_\theta(\cdot)$. This network consists of two main components: a deep feature extraction (DFE) module and an image generation (IG) module.

Initially, the input list-mode data, $\hat{Y}$, is padded with zeros or cropped to a size of $4 \times 4096$ and is fed into a 1d convolution layer which is designed to map the input 4-channel events to a higher-dimensional feature space. The convolution layer utilizes a kernel size of 1 and the output dimension of 180 which means each event is separately expanded to 180 dimensions by using a shared convolution kernel. Next, the expanded list data is input to the DFE module to generate a 256-dimensional deep feature. Finally, the deep feature is fed into the IG module to obtain the final reconstructed image $\hat{X}$.

As seen in Fig.2, the main body of the DFE module is 4 residual Swin Transformer blocks (RSTB) which is modified to be adapted for 1-D input. An RSTB is composed of 4 Swin transformer (ST) layers, coupled with a linear layer and a residual connection. The ST layer employs Layer Normalization (LayerNorm), window multi-head self-attention (W-MSA) or shifted window multi-head self-attention (SW-MSA), and multi-layer perceptron (MLP) layers as the detail structure shows. As the CC reconstruction need more than one CC event to determine the orientation of the source, the window used to calculate the W-MSA/SW-MAS in the local region only slides along the event direction to extract features exhibited by adjacent events. From our previous experience, the window size is set to 8 because 8 effective events are sufficient for single source reconstruction. In addition to the main body, the DFE module utilizes a residual connection, a 1-D convolutional layer, and a linear layer. The convolutional layer and the linear layer are designed to integrate global features.

The IG module upsamples the deep feature through a cascade of transposed convolution layers and Leaky Rectified Linear Units (ReLU). The $1 \times 1 \times 256$ deep feature is upsampled into $256 \times 256 \times 256$ feature maps through 7 steps and finally merged into $256 \times 256 \times 1$ output $\hat{X}$ through a 2-D convolutional layer.

### B. The Loss Function

Different from conventional radiation imaging like Computed tomography (CT) and Single-Photon Emission Computed Tomography (SPECT), applications of CC imaging tend to achieve source detection and localization rather than complicated radioactive distribution reconstruction. The main reason is that CC imaging is based on list-mode events whose acquisition exhibits significant stochastic characteristics. Therefore, in order to constrain the optimization of SwinCCIR for better practical applications, we define a hybrid loss to enhance its ability of source detection and localization. First, we adopt a weighted mean square error (MSE) as the shape loss to constrain the shapes and locations of the reconstructed sources. The shape loss is defined as follow.

$$L_1 = \frac{1}{J}[w_1 \sum_j^{X_j > t}(X_j - \hat{X}_j)^2 + w_2 \sum_j^{X_j < t}(X_j - \hat{X}_j)^2] \quad (5)$$

Where $w_1$ and $w_2$ are weight parameters, $t$ is the threshold used to divide the region where the source is located. This loss aims to overcome the deformation of the source shape caused by directly using MSE in the case of sparse labels. Then, as the multi-interaction events might contain false coincidence events and noise, the relative source intensity might be inaccurate in the label. To enhance the stability of training and the detection capability of this model, we utilize the dice loss which is commonly used in segmentation tasks. The estimated source distribution $\hat{X}$ and the real distribution $X$ are converted into binary images $\hat{X}_b$ and $X_b$, through the same threshold $t$, the dice loss is then defined as below.

$$L_2 = 1 - \frac{2|\hat{X}_b \cap X_b|}{|X_b| + |\hat{X}_b|} \quad (6)$$

Thus, the overall loss function of SwinCCIR can be expressed as:

$$\mathcal{L}(\hat{X}, X) = \alpha L_1 + \beta L_2 \quad (7)$$

where $\alpha$ and $\beta$ are hyperparameters used to adjust the loss ratio.

## IV. EXPERIMENTAL STUDY

### A. Datasets

In this study, two datasets are used to complete the training and testing of the model. One is the simulation dataset and the other is practical datasets measured by the CC prototype introduced in Section II-B. The details of the two datasets are described below:

1) Simulation dataset (high quality): This dataset is made through Monte Carlo simulation using a simulation toolkit Geant4[32,33,34]. We model the two-layer CC and radioactive sources in Geant4 based on the practical CC prototype. We also implement the characteristics of our prototype in the simulation such as the intrinsic position resolution and the energy resolution. The detector in simulation is pixelated and accurate interaction positions are replaced by the center of the nearest pixel. Besides, interaction positions are recorded as the centroids weighted by deposited energies which are the same as the practical output of the detector. Subsequently, random noise that confirms to Gaussian distribution whose standard deviation is determined with reference to the energy resolution of our CC prototype is added to the deposited energies according to the energy, ensuring the simulation results reflect practical conditions. A positron source $^{22}$Na is selected as the simulation gamma source because it has two characteristic gamma energies namely 511 keV and 1275 keV which are closed to that of gamma sources used in medical and industrial applications. The simulation repeats for 400 times with the source placed in different places. Besides, to enlarge the dataset, we divide the multi-interaction events in each simulation into 25 subsets according to positions of the interactions in the first layer, i.e., merging adjacent 2×2 pixels as an individual scatterer. After filtering samples with few events, this dataset contains 9236 single-source samples, of which the sample with the fewest events only contains 67 events and the sample with the most events contains 14462 events. Labels are modeled as the spherical Gaussian distributions whose peaks locate at the relative orientation of sources to the scatterer centers and standard deviations are set as the angular resolution of the CC prototype. Each label is a 256 × 256 double matrix, with the horizontal index being the azimuth angle evenly divided from 0 to π and the vertical index being the evenly divided cosine value of the polar angle from 1 to -1. The image range of labels is 0-1.

2) Practical dataset (low quality): This dataset is made through practical measurement using our CC prototype. Same as the simulation, an approximately 9 MBq point-like $^{22}$Na source is used as the gamma source. Measurement repeats for 48 times with the source placed in different places and each measurement lasts 1 hour. This dataset contains 1500 single-source samples, of which the sample with the fewest events contains 127 events and the sample with the most events contains 750 events. The format and generation of labels are the same as those of the simulated dataset. As both the detector and the source are sealed, their absolute positions cannot be precisely determined; in this experiment, only the relative movement of the source can be accurately measured. As shown in Fig.3, the CC is fixed on the experiment platform and shielded with lead blocks. The $^{22}$Na source is placed on the slide table whose movement precision is better than 1 mm.

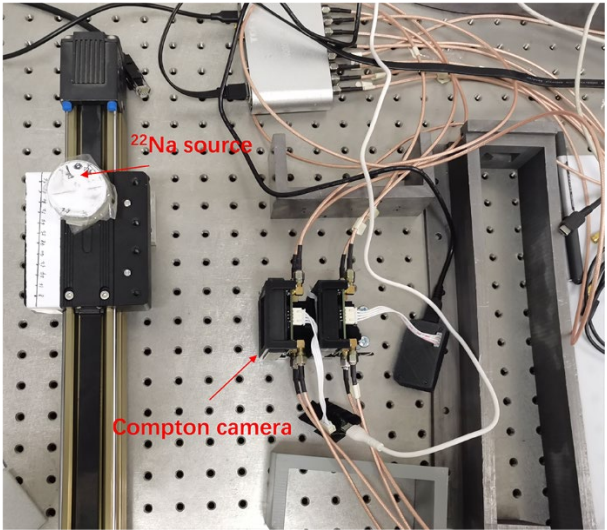

Fig. 3. The picture from the experiment of the collection of the practical dataset.

### B. Experimental Setup

Data augmentation: As CC imaging is based on the list-mode events, we conduct data augmentation from two perspectives. First, we add noises to the positions and energies of interactions. Based on the signal acquisition accuracy of our CC, we added uniformly distributed noise to the interaction positions and Gaussian-distributed noise to the energies.

Second, we randomly select events from the sample as input when training, given that CC imaging is indifferent to the sequence of events and only needs a certain number of events. Furthermore, as the acquisition of events of CC is time independent, the combination of events in different measurements is equivalent to the results of measuring sources at different places simultaneously. We generate multi-source samples by merging two or three original samples to test the imaging capability of multiple sources of the SwinCIR. In the simulation dataset, 10000 double-sources samples and 10000 triple-sources samples is generated randomly, while 2059 double-sources samples and 1542 triple-sources samples is added in the practical dataset. In addition, image mirroring is applied during the training process and the relative positions in the list-mode data are changed accordingly.

Quantitative evaluation metrics: The quality of reconstructed images is evaluated using two metrics namely peak signal-to-noise ratio (PSNR) in dB and structural similarity (SSIM), which is defined below:

$$\text{PSNR} = 20 \cdot \log_{10}\left(\frac{\max{(X, \hat{X})}}{\sqrt{\frac{1}{J}\Sigma_j^J (X_j - \hat{X}_j)^2}}\right) \tag{8}$$

$$\text{SSIM} = \frac{(2\mu_X \mu_{\hat{X}} + C_1)(2\sigma_{X\hat{X}} + C_2)}{(\mu_X^2 + \mu_{\hat{X}}^2 + C_1)(\sigma_X^2 + \sigma_{\hat{X}}^2 + C_2)} \tag{9}$$

where $\mu_X$ and $\mu_{\hat{X}}$ are the average values of $X$ and $\hat{X}$, $\sigma_X$ and $\sigma_{\hat{X}}$ are the standard deviation of $X$ and $\hat{X}$, $\sigma_{X\hat{X}}$ is the covariance between the $X$ and $\hat{X}$. $C_1$ and $C_2$ are constants to avoid instability when the SSIM is closed to zero. Empirically, $C_1$ and $C_2$ are set to 1e-4 and 1e-3.

Training settings: Approximately 80% samples in each dataset are used for training and the rest 20% for validation and testing. The model learns with the Adam optimizer at a learning rate of 2e-4. Multi-step strategy is applied to the learning rate which will decrease by a factor of 2 every 40000 steps. For the loss function, $\alpha$ and $\beta$ are set to $\frac{1}{|L_1|}$ and $\frac{1}{|L_2|}$. $w_1$ is set to 1 and $w_2$ is set to 0.2. The threshold $t$ that used in weighted MSE and image binarization is set to 0.02. Training takes approximately 24 hours on an NVIDIA RTX 3090 GPU with a power limit of 225 W.

Reference methods: To prove the effectiveness of SwinCCIR, the Comptonnet from recent literature and 3 other conventional reconstruction algorithms are used for comparison. When reproducing the Comptonnet, we adjusted the input size to be the same as that of the SwinCCIR. The loss of the Comptonnet is MSE which is the same as that in [26]. Adam optimizer and multi-step strategy are applied. Three reconstruction algorithms are SBP, MLEM and SOE as mentioned in Section I. Before the conventional algorithms are performed, the events are filtered and reordered according to the energies of interactions.

### C. Experimental Results

Fig. 4 shows the visual quality of CC imaging using all 5 methods under consideration for single-source image, double-sources image and triple-sources image in the simulation dataset. For each kind of source images, two samples, one with fewer events and the other with more events, were selected to show the performance of each method. For the two deep learning methods, all the test samples are zero-padded to integer multiples of the input length and divided into several patches. The final output is then obtained by summing the output weighted by the number of valid events in each patch.

The deep learning methods demonstrate a significant enhancement compared to conventional methods. The results of three conventional methods can only roughly discern just the number and locations of sources. The SBP method is acutely affected by annular artifacts originated from the fundamental imaging principle and exhibit the worst performance, whereas the SOE method which is based on Markov chain suffers from a pronounced decline in imaging accuracy due to the lack of events. As for the current benchmark of iterative reconstruction algorithms, MLEM, the shapes of the reconstructed sources still undergo considerable deformation though it exhibits relatively excellent performance, leading to difficulties in accurate localization of sources. Results of the Comptonnet exhibits artifacts of false sources for multi-source distribution, while the SwinCCIR method outperforms Comptonnet in the accuracy of the number of sources and their relative intensities.

Besides, Table I presents the average quantitative performance of the 5 methods on three kinds of simulation images in the test set. The SwinCCIR yields performance superior to that of the other techniques for all three kinds of source distributions.

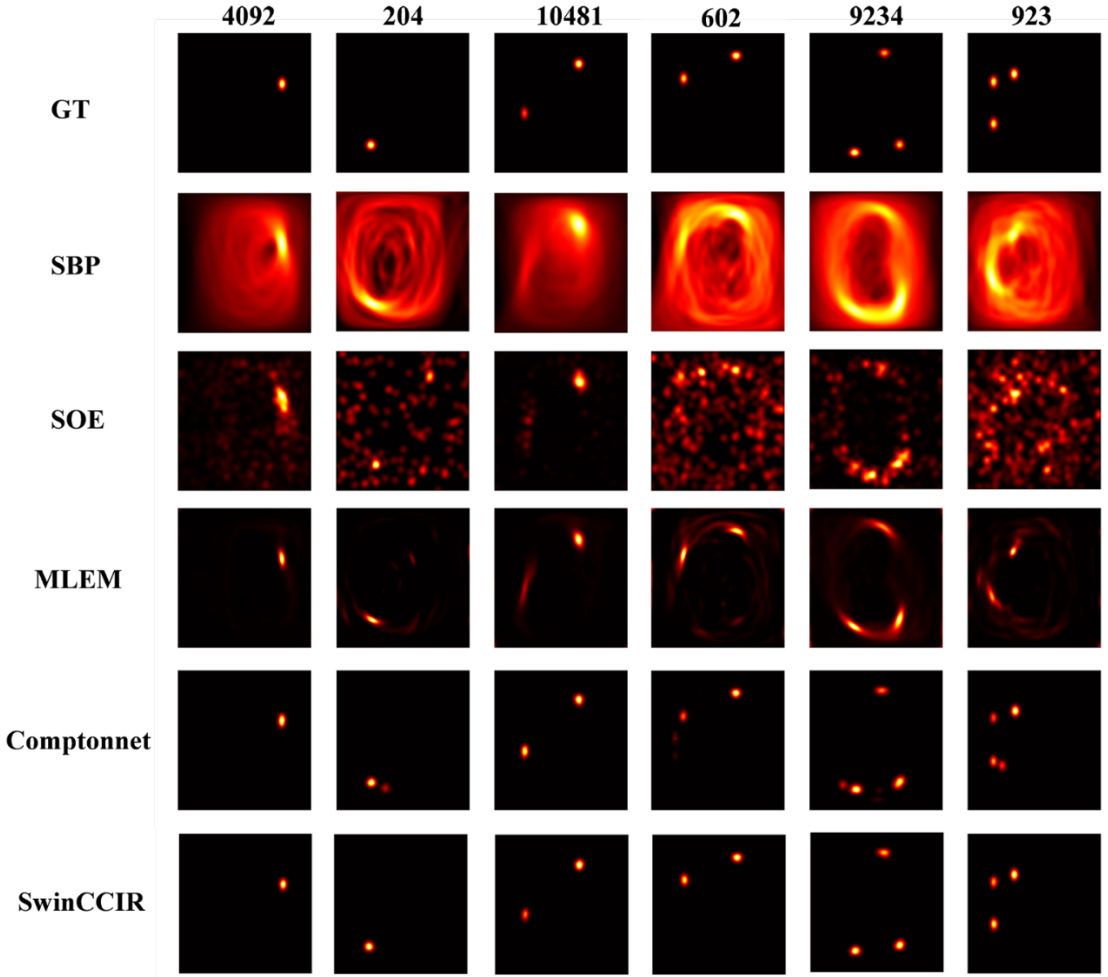

Fig. 4. Visual quality for the 6 selected source distribution images in the simulation test set. The number at the top of each column represents the total number of events contained within that sample.

Fig. 5 and Table II present the results and evaluation metrics of six testing images from the practical dataset.

Despite a decline in performance across all methods due to deteriorating dataset quality, SwinCCIR continued to demonstrate favourable performance. It notably outperforms Comptonnet in reconstructing both the shape and relative intensity of the source.

TABLE I
QUANTITATIVE PERFORMANCE ON THE SIMULATION TEST SET

|  | Single source | | Double sources | | Triple sources | |
|---|---|---|---|---|---|---|
|  | PSNR | SSIM | PSNR | SSIM | PSNR | SSIM |
| SBP | 11.1631 | 0.0110 | 9.4354 | 0.0085 | 8.0542 | 0.0095 |
| SOE | 17.2900 | 0.1970 | 16.5079 | 0.1055 | 15.3357 | 0.05677 |
| MLEM | 31.5929 | 0.7507 | 28.4798 | 0.5991 | 26.3865 | 0.5016 |
| Comptonnet | 46.5231 | 0.9964 | 38.9882 | 0.9894 | 34.3255 | 0.9791 |
| SwinCCIR | **49.6123** | **0.9989** | **41.9982** | **0.9949** | **37.1544** | **0.9876** |

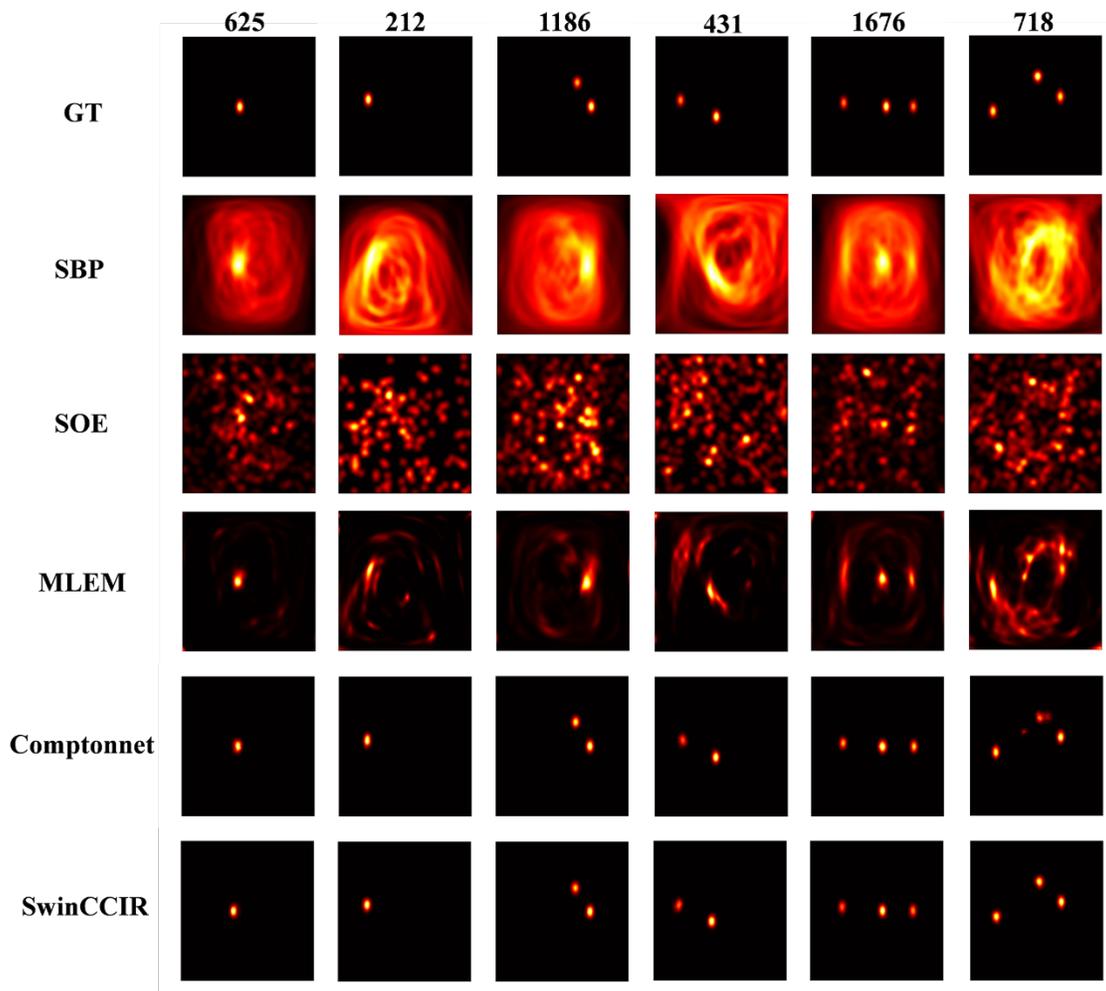

Fig. 5. Visual quality for the 6 selected source distribution images in the practical test set. The number at the top of each column represents the total number of events contained within that sample.

Furthermore, to assess the requirements of the number of events to achieve accurate reconstruction, we generate a sequence of new samples with varying numbers of events and then reconstruct the images using methods mentioned above. Based on previous results, multiple-sources imaging proves more challenging than single-source. Therefore, to better demonstrate the superiority of deep learning methods, this experiment concentrates on triple-source samples. Specifically, these samples are composed of an equal number of events randomly selected from 3 single-source samples within the simulated dataset; consequently, the intensities of the 3 sources in the corresponding label are also considered equal. As the SOE algorithm requires sufficient events to achieve high-quality reconstruction, and the objective of this experiment was to determine the minimum number of events required by different methods, the SOE algorithm was not included in the comparison. The results are shown in Fig. 6. Again, the SwinCIR demonstrates optimal performance that achieves accurate localization of all three sources using only 100 events (300 in total) in each single source sample, while the Comptonnet requires 200 events per sample to localize the third source. As for the MLEM, severe artifacts emerge with few events, rendering source localization virtually impossible. The artifacts are suppressed only until over 300 events per sample were used in reconstruction. Besides, Fig.7 shows the trends of two metrics as the number of input events varies where only the two deep learning methods are compared due to their tremendous performance. These results reveal that the measurement process of the CC can be shortened by using the SwinCCIR for reconstruction.

TABLE. II
QUANTITATIVE PERFORMANCE ON THE PRACTICAL TEST SET

| | Single source | | Double sources | | Triple sources | |
|---|---|---|---|---|---|---|
| | PSNR | SSIM | PSNR | SSIM | PSNR | SSIM |
| SBP | 10.1339 | 0.0173 | 8.1290 | 0.0066 | 6.9717 | 0.0064 |
| SOE | 15.3002 | 0.2071 | 14.2448 | 0.0752 | 13.6605 | 0.0416 |
| MLEM | 28.2504 | 0.7281 | 26.2840 | 0.5407 | 22.9867 | 0.4157 |
| Comptonnet | 41.5080 | 0.9949 | 37.7405 | 0.9905 | 32.5693 | 0.9805 |
| SwinCCIR | **48.9452** | **0.9980** | **41.7222** | **0.9944** | **36.0055** | **0.9866** |

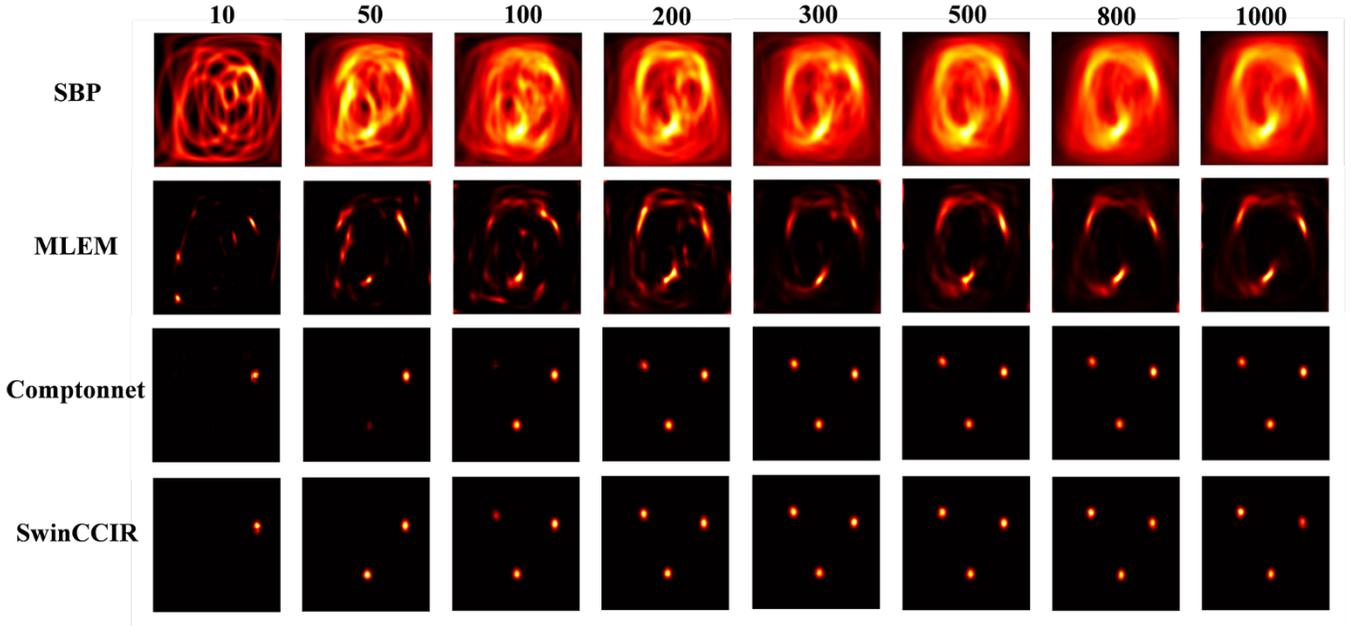

Fig. 6. Visual quality for the same source distribution reconstructed using different numbers of events per sample. The number at the top of each column represents the number of events selected from each sample.

## D. Ablation study

We now conduct a series of ablation experiment to evaluate the contribution of several components in our SwinCCIR framework. First, to evaluate the impact of the hybrid loss function, we apply the hybrid loss function to the Comptonnet. Second, we replace the two MLP layer in the Comptonnet with our DFE module and contain the hybrid loss function.

Third, we replace the convolution layers in the Comptonnet with the IG module and train also with the hybrid loss. As comparison, the three experiments are compared to the origin Comptonnet and SwinCCIR on the simulation dataset. The quantitative evaluation is shown in Table III. The results show that the DFE module provides significant improvement on the accuracy of the CC reconstruction and the hybrid loss contribute to enhancing training efficacy in this category of tasks.

TABLE. III
ABLATION PERFORMANCE

| Hybrid loss | DFE module | IG module | PSNR(dB) |
|:---:|:---:|:---:|:---:|
| × | × | × | 39.7021 |
| √ | × | × | 40.8126 |
| √ | √ | × | 41.7233 |
| √ | × | √ | 40.9976 |
| √ | √ | √ | **42.2662** |

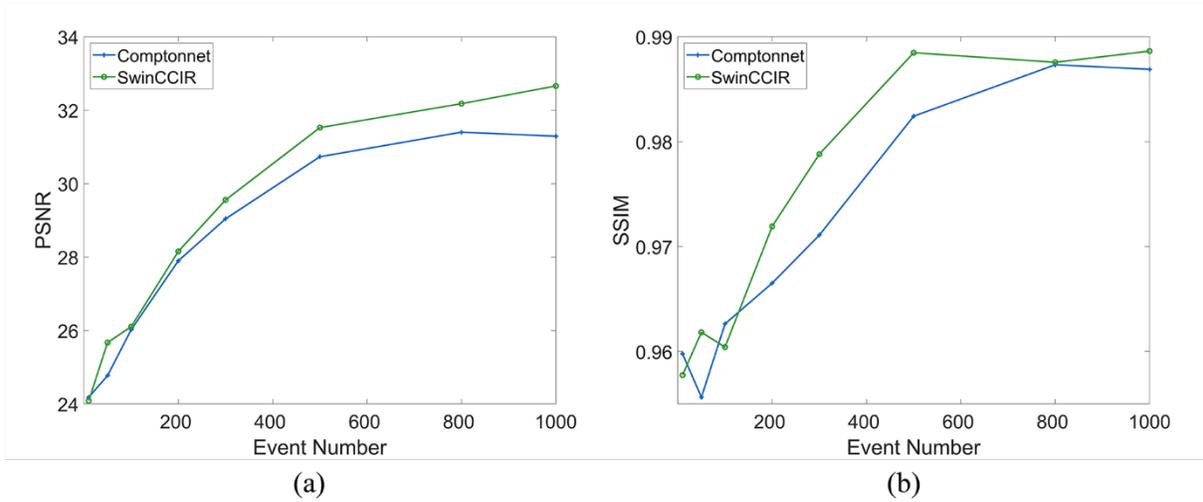

Fig. 7. The quantitative performance of the two deep-learning methods on samples with different numbers of events. (a) is the PSNR and (b) is the SSIM.

## V. CONCLUSION

In this paper, we propose an end-to-end Compton camera imaging framework called SwinCCIR aiming at solving the problems of low quality of multi-interaction events and discontinuity between the event and image domain in Compton camera imaging. The SwinCCIR consists of a deep feature extraction module and image generation module. The deep feature extraction module employs 1-D residual Swin transformers and attention mechanisms to capture the deep relationships between Compton scattering events. Unlike existing neural field methods that focus on either event or image domains, our approach directly achieves Compton camera image reconstruction. The framework is test on both simulation data and practical data. Compared to several commonly used iterative reconstruction algorithms in Compton imaging, the quality of the reconstruction results has been significantly improved. This article explores the possibility of implicitly expressing the reconstruction system matrix using a small subset of measurement data, bypassing the negative effects of the inherent systematic errors, which is expected to highly improve the reconstruction efficiency and quality of Compton cameras.